# Low-rank based motion correction followed by automatic frame selection in DT-CMR

Fanwen Wang[1,2], Pedro F.Ferreira[2,3], Camila Munoz[2,3], Ke Wen[2,3], Yaqing Luo[2,3], Jiahao Huang[1,2], Yinzhe Wu[1,2], Dudley J.Pennell[2,3], Andrew D. Scott[2,3], Sonia Nielles-Vallespin[2,3] and Guang Yang[1,2]

1. Bioengineering Department and Imperial-X, Imperial College London, London W12 7SL, UK;
2. Cardiovascular Research Centre, Royal Brompton Hospital, London SW3 6NP, UK;
3. National Heart and Lung Institute, Imperial College London, London SW7 2AZ, UK;

## Introduction

Diffusion tensor-based cardiovascular magnetic resonance imaging (DT-CMR) offers an innovative way to visualize in-vivo cardiomyocytes' orientation. Equipped by breath-holds and cardiac triggers to mitigate motion during its extensive acquisition process, which includes multiple diffusion directions and averages, DT-CMR is still susceptible to motion artifacts. Conventionally, pairwise rigid/affine registrations are employed for correcting in-plane motion in STEAM (Stimulated Echo Acquisition Mode) and SE (Spin Echo) data, and frames exhibiting signal loss or through-plane motion are manually excluded during post-processing. However, the selection of an appropriate reference image for registration is particularly challenging due to the diverse contrasts and the influence of cardiac and respiratory motion, which influences the successive pairwise registration. The subjective nature of excluding images with poor registration, through-plane motion, or low signal-to-noise ratio (SNR) presents a further challenge. Hence, we adopted the use of low-rank images, which facilitated quicker registration, coupled with an automatic frame selection process to enhance the robustness of the DT-CMR post-processing workflow.

## Methods

12 healthy volunteers were scanned at both 3T and 1.5T, with a STEAM and SE-EPI with spatial resolution $2.8\times2.8\times8$ mm$^3$, *b0* data and 6 encoding directions at *b* = 150 and 600 s/mm$^2$ with more than 9 averages in total, in short-axis view at end-systole and end-diastole frames. All the frames were acquired with matrix size of 256×96 and the central region were automatic cropped as 96×96. Some SE-diastole cases were discarded due to the quality control.

We conducted a comparative analysis between the proposed method and an in-house post-processing workflow (Figure 1). The in-house method was characterized by manual circular cropping aimed at diminishing the high-intensity signal from adjacent organs, a rapid rigid registration (1), and a manual selection of frames. We expressed all the acquired images $A = N_x \times N_y \times N_{ave} \times N_{DWI}$ in the Tucker form (2) :

$$A = \Phi_{xy} \cdot \Phi_{dyn}$$

where $\Phi_{xy}^{(N_x \times N_y \times L_1)}$ is the image basis matrix and $\Phi_{dyn}^{(L_1 \times N_{ave} \times N_{DWI})}$ is the dynamic factor tensor. $L_i$ is the $i^{th}$ rank for a given basis matrix. After automatic central cropping, the geometric average $A^{L1=1}$ images were taken as the reference since they disentangled the respiratory and local shape deformations (Figure 2). Derived from a PCA denoised method (3), we set $A^{L1=6}$ as moving frames to suppress noise. Denoised images $A^{L1=6}$ replaced manual circular cropping, enhancing registration robustness and expediting convergence. Affine/rigid transformation parameters, derived from the registration between the averaged and denoised frames, were applied to the original diffusion-encoded frames for improved accuracy.

The region of interest (ROI) was determined to be a donut-shaped area centred on the manually annotated blood pool for automatic frame selection. The inner radius was set at 95% of the shortest distance from the labelled myocardium boundary, while the outer radius extended to 105% of the maximum distance, ensuring precise delineation of the myocardial tissue. We employed a Pearson correlation to identify and excluded outliers situated more than three scaled median absolute deviations below the median value (4).

The RMSE, R-square of the fitting of line profile of helix angle (HA), the number of negative eigenvalues in the myocardium per mile, and horizontal and vertical visualization were chosen as evaluation (5).

Both affine/rigid registrations for SE/STEAM were implemented using itk-elastix (6) in python. The previous rigid registration used dft-registration (1) in matlab. All the experiments were implemented on a Linux server with 64-core CPU.

## Results

The proposed method showed HA line profiles with higher R-square, comparable visualization with less manual burden (Figure 3). It's worth noting that the frame selection method was previously applied on STEAM-EPI. We found the automatic rejection also generalized on SE data with similar performance as manual selection. For dft-registration, although it enjoyed high computational efficiency, it failed if suboptimal manual cropping included the high intensity of the surrounding chest wall (Figure 4).

The statistical metrics were calculated for SE and STEAM respectively. The proposed registration outperformed the in-house workflow which just relied on the original images. It reached higher R-square, lower RMSE and less percentage of negative eigenvalues in the myocardium in both SE and STEAM. (Table 1)

## Conclusion and Discussion

We proposed a method which used low-rank decomposition as pre-processing to enable a robust rigid/affine registration. Automatic frame selection was guided by Pearson correlation within the myocardium-segmentation guided part. Instead of choosing the reference image with brightest frame, the proposed method averaged the low-rank frames which discards the local deformation, respiratory motion and noise as the reference to boost the registration. The proposed method encountered limitations where automatic cropping struggled when the heart was not centrally located, requiring manual adjustment. Additionally, despite the dft-registration's high efficiency, processing 63 slices in approximately 0.2 seconds, both rigid and affine registrations took around 20 seconds.

# Figures

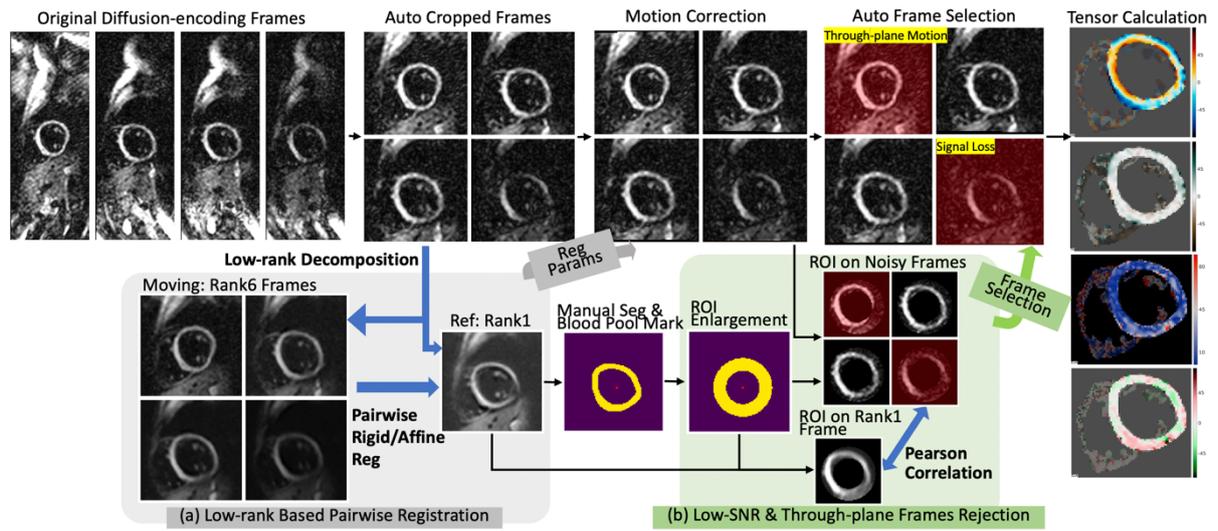

Figure 1 The semi-automatic pipeline of the proposed low-rank registration with automatic frame selection. The reference was set as the geometric average of rank-1 image disentangles local heart deformation and respiratory motion. Manual segmentation of myocardium and labeling of the center of blood pool was utilized to guide the frame selection.

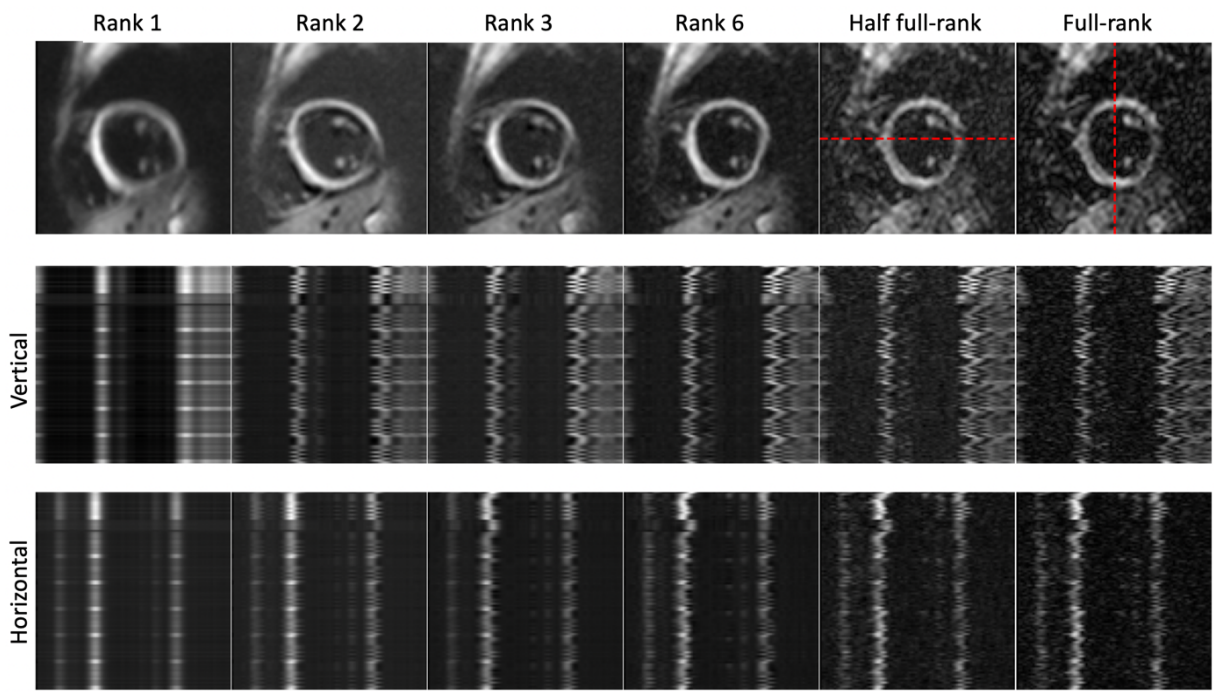

Figure 2 The evolving ranks of a frame with comparatively image stack visualization. As rank gradually increases, the image includes respiratory shift, local deformation, diverse diffusion contrasts and noise.

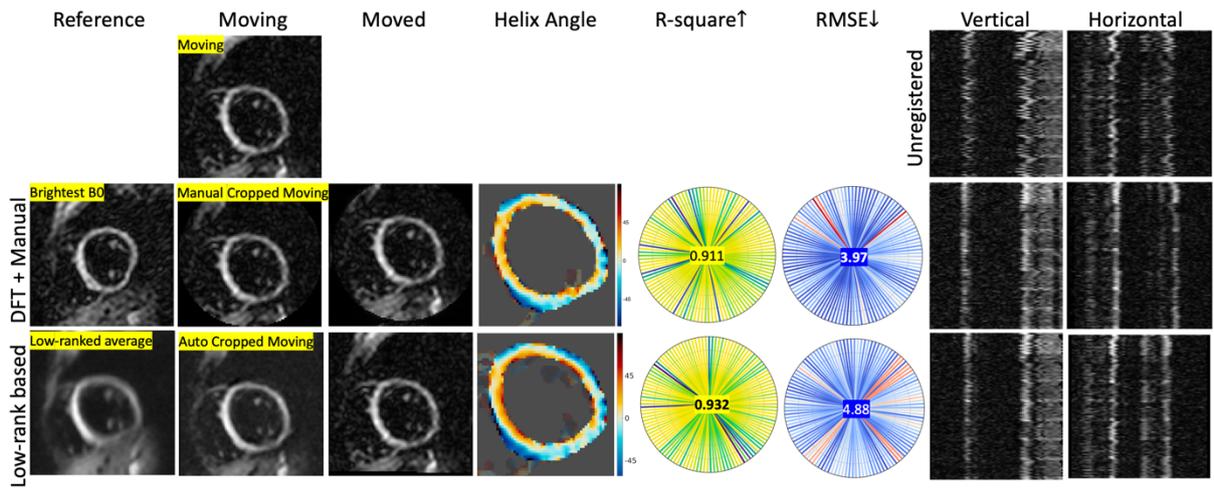

Figure 3 A representative case was processed by the previous workflow and the proposed low-rank based registration. DFT+Manual is the previous in-house post-processing pipeline that uses rigid registration followed by manual frame selection. Low-rank based pipeline is the one we proposed with automatic frame selection. R-square and RMSE quantify the fitting of the helix angle.

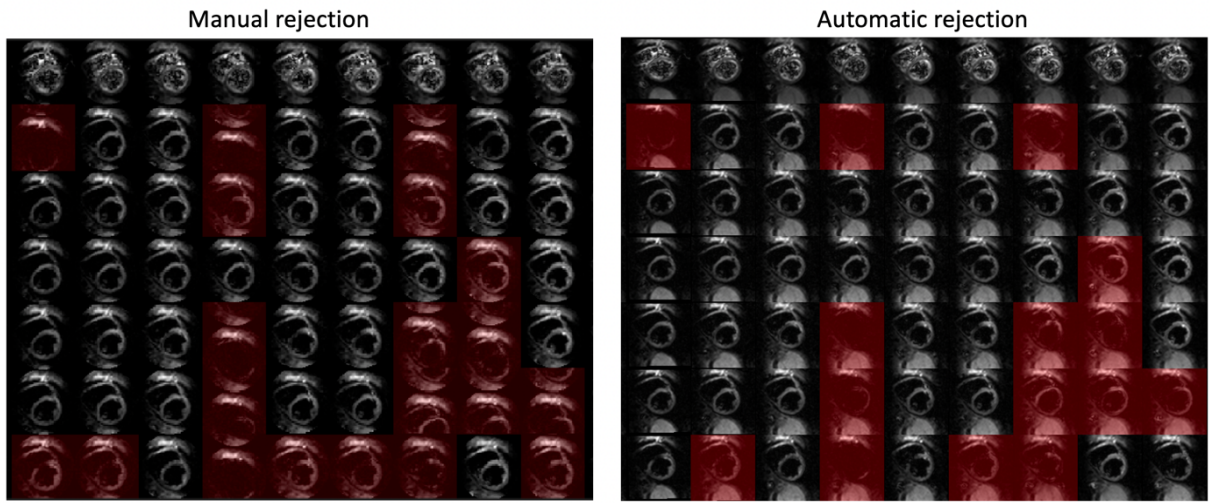

Figure 4 Manual frame selection and automatic frame selection on one case. Some frames with misregistration need to be further discarded when using dft-registration.

|  |  | R-square↑ | RMSE↓ | Nega1(‰)↓ | Nega2(‰)↓ |
|---|---|---|---|---|---|
| SE | DFT + manual | 0.901±0.116 | 6.360±4.681 | 1.14±0.03 | 0.00±0.01 |
|  | Low rank + manual | **0.914±0.113** | 5.946±4.054 | 0.49±0.02 | 0.00±0.00 |
|  | Low rank + auto | 0.911±0.115 | **5.898±4.149** | **0.30±0.02** | **0.00±0.00** |
| STEAM | DFT + manual | 0.914±0.108 | 5.134±3.537 | 11.3±0.41 | 0.00±0.01 |
|  | Low rank + manual | 0.924±0.099 | 4.486±3.159 | 5.21±0.22 | 0.00±0.00 |
|  | Low rank + auto | **0.925±0.096** | **4.467±3.120** | **5.10±0.21** | **0.00±0.00** |

*Table 1 Statistical metrics of the proposed method. DFT and low-rank denote the rigid registration and the proposed method. The R-square and RMSE is shown in mean ± std. Nega X stand for the per hundred of pixels with X negative eigenvalues among all the pixels in the myocardium. They are shown in median ± interquartile. Manual and auto stands for the manual and automatic frame selection respectively.*